\documentclass[creativecommons]{eptcs}

\usepackage{iftex}

\usepackage{listings}
\usepackage{xcolor}
\usepackage{graphicx}
\usepackage{appendix}

\ifpdf
  \usepackage{underscore}         
  \usepackage[T1]{fontenc}        
\else
  \usepackage{breakurl}           
\fi

\title{Verification Challenges in Sparse Matrix Vector Multiplication in High Performance Computing: Part I}
\author{Junchao Zhang
\institute{Argonne National Laboratory\\Illinois, USA}
\email{jczhang@anl.gov}
}
\def\titlerunning{Verification challenges in SpMV in HPC: Part I}
\def\authorrunning{J. Zhang}
\begin{document}
\maketitle

\begin{abstract}
Sparse matrix vector multiplication (SpMV) is a fundamental kernel in scientific codes
that rely on iterative solvers. In this first part of our work,
we present both a sequential and a basic MPI parallel implementations of
SpMV, aiming to provide a challenge problem for the scientific software verification community.
The implementations are described in the context of the PETSc library.

\end{abstract}

\section{Introduction}
\begin{figure}[h]
  \centering
  \includegraphics[width=0.8\textwidth]{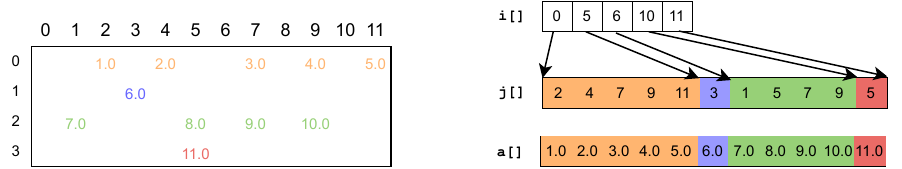}
  \caption{{\it Left}: A 4 $\times$ 12 sparse matrix. {\it Right}:  its compressed sparse row (CSR) representation.}
  \label{fig:seqcsr}
\end{figure}

Solving sparse linear systems of $Ax=b$ lies in the heart of scientific computing.
Iterative Krylov subspace methods \cite{saad2003iterative} are a key algorithm for that,
especially for large scale computations.
SpMV is a crucial kernel in these methods.
The popular math library, PETSc (the Portable, Extensible Toolkit for Scientific Computation)\cite{petsc-user-ref}, contains
a suite of Krylov methods and SpMV implementations.
Since the matrix is usually very sparse,
it is wise to only store nonzeros to save memory.
The compressed sparse row (CSR), known as {\tt MATAIJ} in PETSc,
is a commonly used sparse matrix storage format.
It represents an {\tt m} $\times$ {\tt n} matrix with total {\tt nnz} nozeros in three arrays: {\tt a[nnz]}, {\tt i[m]} and {\tt j[nnz]},
 where {\tt a[]} and {\tt j[]} store the nonzeros and their column indices row-wisely, respectively, and {\tt i[]}
stores indices pointing to the start of each row in {\tt a[]} and {\tt j[]}.
Apparently, {\tt i[0] = 0} and row {\tt k} has {\tt i[k+1] - i[k]} nonzeros. {\tt i[m]} points to the next place
after the last nonzero. Figure \ref{fig:seqcsr} shows a CSR example for a 4 $\times$ 12 sparse matrix.

Matrix vector multplication has a very simple definition.
Denote $A = \{a_{ij}\}$, $x = \{x_j\}$, $y = \{y_i\}$, then $y = Ax$ is defined by
$y_i = \sum_{j=0}^{n-1}a_{ij} x_j$, with $0\leq i < m$, $0 \leq j < n$.
However, it is the storage format, data distribution, parallelization and optimizations
that make implementations complicated.
Among PETSc's many SpMV implementations,
there is an MPI parallel one with sophiscated optimizations.
We won't elaborate that one here and will leave it to part II of the work in the future.
In this paper (part I), we present a sequential and a basic MPI implementations, which are simple but still share good
ingredients with the optimized MPI one.
We first describle the structure of the code, then the two implementations and the verification challenges within, and conclude at the end.

\section{Code structure, input and output}
We provide standalone implementations written in the C language in a
public repository\footnote{\url{https://github.com/jczhang07/cs2}}.
One only needs to specify a C or MPI compiler to compile them.
The input data is hardwired in source code with global varibles.
We provide matrix $A$ of size {\tt M} $\times$ {\tt N} and {\tt NNZ} nonzeros in three arrays
{\tt Gi[M+1]}, {\tt Gj[NNZ]}, {\tt Ga[NNZ]},
vector $x$ in {\tt Xa[N]}, and the correct answer $z = Ax$ in {\tt Za[M]}. There is also an array {\tt Ya[M]}
to provide space for vector $y$, which will store the computed result of $Ax$. The code will compute the square of
the 2-norm of $|| y - z ||$, which should be zero, and return a non-zero code if $y$ is wrong.
The input data was generated by the accompanied Python script csr.py. One can modify parameters in the script and re-run
it to generate a different set of inputs.

\section{Sequential implementation}
\label{sec:seq}
In this implementation (seq.c), marix $A$, vectors $x$, $y$ and $z$ reside in one process.
We directly build $A$, $x$, $y$ and $z$ from the global arrays mentioned above.
Sizes of $x$, $y$ and $z$ must conform to $A$'s shape.
\vspace{-10pt}
\begin{figure}[h]
\begin{minipage}{0.35\textwidth}
\begin{lstlisting}[language=C]
typedef struct {
  int     m, n; // size
  int    *i, *j;
  double *a;    // values
} Mat;

typedef struct {
  int     n; // size
  double *a; // values
} Vec;
\end{lstlisting}
\end{minipage}
\hfill
\begin{minipage}{0.55\textwidth}
  \begin{lstlisting}[language=C]
int i, j;
for (i = 0; i < A.m; i++) {
  y.a[i] = 0.0;
  for (j = A.i[i]; j < A.i[i + 1]; j++)
    y.a[i] += A.a[j] * x.a[A.j[j]];
}
\end{lstlisting}
\end{minipage}
\vspace{-10pt}
\caption{{\it Left}: sequential matrix and vector types. {\it Right}: sequential SpMV kernel.}
\label{fig:seqspmv}
\vspace{-5pt}
\end{figure}

Figure \ref{fig:seqspmv} shows the sequential matrix and vector types and the SpMV kernel.
One needs to verify the code computes $y=Ax$. In other words, assume those $a_{ij}$
not in $A$'s CSR representation are zero, then $y_i$ = $\sum_{j=0}^{n-1}a_{ij} x_j$,
for $0\leq i < m$, $0 \leq j < n$, without counting floating point round-off errors.


\section{Basic MPI implementation}
\label{sec:mpibasic}
This implementation (mpibasic.c) uses MPI parallelsim.
The global matrix is block-distributed by row. Each MPI process has a piece (submatrix) of the
matrix, where the row size of the submatrix can be set by users
or automatically computed by PETSc. We call it the matrix row layout.
Though matrix columns are not distrubuted, they also have a layout that can be set by users.
These row and column layouts are properties of the matrix, by which we distribute vectors $y$ and $x$ respectively.
In PETSc, one can use {\tt MatCreateVecs(A, \&x, \&y)} to create
vectors with conforming size and layout suitable for doing $y = Ax$.
We use {\tt M}, {\tt N} to represent the global size of the matrix,
and {\tt m}, {\tt n} for the local size of the diagonal block of the matrix
on this process. {\tt m}, {\tt n} are also the local size of $y$ and $x$ respectively.
In this implementation, we first compute the matrix layouts.
Suppose {\tt size} is the size of {\tt MPI_COMM_WORLD} and {\tt rank} is the rank of the current MPI process,
{\tt m}, {\tt n} are computed as \verb|m = M/size + (M%size > rank ? 1 : 0)|,
\verb|n = N/size + (N%size > rank ? 1 : 0)|. This is the formula PETSc would use
if users don't set {\tt m} and {\tt n}.
With the layouts set,
each process knows the indices of its first row {\tt rstart} and first column {\tt cstart}.
Then it pulls out its data from the global arrays.
{\tt A.j[]} and {\tt A.a[]} can share the space with {\tt Gi[]} and {\tt Ga[]}, but we
have to allocate {\tt A.i[]} and populate it with shifted indices from {\tt Gi[]}.
Similarily, we can build $x$, $y$ and $z$ by pointing them to
{\tt Xa[cstart]}, {\tt Ya[rstart]} and {\tt Za[rstart]} repectively.

\begin{figure}[h]
\begin{minipage}{0.35\textwidth}
\begin{lstlisting}[language=C]
typedef struct {
  int     m, n, M, N;
  int     rstart, cstart;
  int    *i, *j;
  double *a;
} Mat;

typedef struct {
  int     n, N;
  double *a;
} Vec;
\end{lstlisting}
\end{minipage}
\hfill
\begin{minipage}{0.60\textwidth}
    \includegraphics[width=1.0\textwidth]{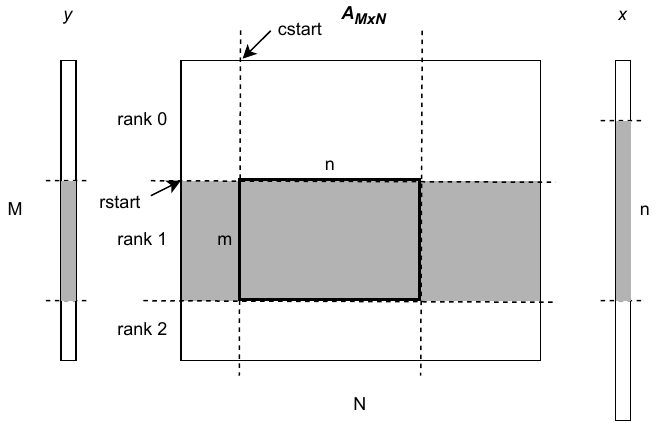}
\end{minipage}
\caption{{\it Left}: MPI parallel matrix and vector types.
{\it Right}: distributed matrix $A$ and vectors $x$ and $y$ on three MPI ranks.
The shadowed parts reside on rank 1.}
\label{fig:mpicsr}
\end{figure}

To multiply the local submatrix $A$ with vector $x$, we need remote entries of $x$. In theory, we only
need entries of $x$ which corresponds to nonzero columns in the local submatrix. But for simplicity,
we gather the whole distributed $x$ to a local vector $X$ of size {\tt N}.
Depending on whether {\tt N} is evenly distributed or not, we can use {\tt MPI_Allgather} or {\tt MPI_Allgatherv}.
With our block distribution formula, this can be tested by {\tt N\%size}. Otherwise,
one has to check whether all {\tt n}'s on processes are equal.
With $X$, we perform a sequential SpMV as in Section \ref{sec:seq} and get the local part of
$y$. After that we compute the partial norm of $||y - z||$, and
then the final norm with help of {\tt MPI_Allreduce}.

For this MPI implementation, one needs to verify 1) the layouts are correct, i.e., $\sum m = M$, $\sum n = N$;
2) the $y$ on each process is the result of a sub-SpMV for $Ax$ where $A$ is the local submatrix and
as if $x$ was not distributed.


\section*{Acknowledgements}
The work is supported in part by the U.S. Department of Energy, Office of Science,
under the FASTMath institute within the Scientific Discovery through Advanced Computing (SciDAC-5) program
under contract DE-AC02-06CH11357.

\nocite{*}
\bibliographystyle{eptcs}
\bibliography{refs.bib}

\appendix
\section{Sequential SpMV code: seq.c}

\begin{lstlisting}[language=C, numbers=left]
// A sequential sparse matrix vector multiplication (SpMV, MatMult in PETSc)
#include <stdio.h>

// For simplicity, use global variables to store an M x N matrix A with NNZ nonzeros in the CSR format,
// and two vectors X, Z with Z = AX. The example was generated by the Python script crs.py
enum {
  M   = 32,
  N   = 36,
  NNZ = 50
};
static int    Gi[M + 1] = {0, 1, 1, 2, 6, 9, 10, 10, 11, 11, 13, 13, 13, 16, 17, 18, 18, 19, 20, 22, 23, 27, 28, 31, 32, 34, 37, 37, 41, 44, 44, 47, 49};
static int    Gj[NNZ]   = {25, 13, 1, 5, 7, 35, 18, 19, 31, 32, 21, 32, 33, 0, 8, 27, 16, 25, 3, 24, 17, 27, 13, 3, 28, 29, 30, 2, 23, 29, 31, 10, 8, 29, 1, 20, 22, 3, 8, 16, 19, 10, 14, 24, 2, 6, 15, 17, 34};
static double Ga[NNZ]   = {8, 3, 7, 5, 6, 7, 1, 9, 8, 9, 9, 9, 5, 1, 5, 8, 4, 4, 2, 11, 3, 8, 9, 7, 7, 4, 2, 2, 7, 6, 9, 3, 4, 2, 2, 9, 7, 4, 7, 8, 1, 8, 6, 1, 3, 3, 6, 6, 1};
static double Xa[N]     = {3, 2, 2, 7, 1, 5, 3, 3, 6, 6, 4, 8, 8, 4, 7, 8, 9, 7, 7, 6, 9, 5, 8, 5, 7, 5, 5, 5, 2, 4, 8, 1, 3, 6, 9, 8};
static double Za[M]     = {40, 0, 12, 113, 69, 27, 0, 45, 0, 57, 0, 0, 73, 36, 20, 0, 14, 77, 61, 36, 95, 4, 68, 12, 32, 141, 0, 148, 81, 0, 63, 51};
static double Ya[M];

// Petsc MATSEQAIJ matrix (aka compressed sparse row (CSR) storage)
typedef struct {
  int     m, n; // local row/column size
  int    *i;    // row pointer
  int    *j;    // column indices
  double *a;    // values
} Mat;

// Petsc VECSEQ vectors
typedef struct {
  int     n; // local size
  double *a; // values
} Vec;

int main(int argc, char **argv)
{
  Mat    A;
  Vec    x, y, z;
  double norm = 0;

  // Build A.  Because this is a sequential code, the local matrix takes all
  A.m = M;
  A.n = N;
  A.i = Gi;
  A.j = Gj;
  A.a = Ga;

  // Build x according to A's column size
  x.n = A.n;
  x.a = Xa;

  // Build y, z according to A's row size
  y.n = A.m;
  y.a = Ya;

  z.n = A.m;
  z.a = Za;

  // Compute y = Ax
  for (int i = 0; i < A.m; i++) {
    y.a[i] = 0.0;
    for (int j = A.i[i]; j < A.i[i + 1]; j++) y.a[i] += A.a[j] * x.a[A.j[j]];
  }

  // Compute the norm of ||y - z||
  for (int i = 0; i < A.m; i++) norm += (y.a[i] - z.a[i]) * (y.a[i] - z.a[i]);

  if (norm > 1e-6) {
    printf("Error in computing y = Ax, with norm = %f\n", norm);
    return 1;
  } else {
    printf("Succeeded in computing y = Ax\n");
    return 0;
  }
}
\end{lstlisting}

\section{Basic MPI SpMV code: mpibasic.c}
\begin{lstlisting}[language=C, numbers=left]
// A basic MPI parallel sparse matrix vector multiplication (SpMV, MatMult in PETSc)
#include <mpi.h>
#include <stdio.h>
#include <stdlib.h>

// For simplicity, use global variables to store an M x N matrix A with NNZ nonzeros in the CSR format,
// and two vectors X, Z with Z = AX. The example was generated by the Python script crs.py
enum {
  M   = 32,
  N   = 36,
  NNZ = 50
};
static int    Gi[M + 1] = {0, 1, 1, 2, 6, 9, 10, 10, 11, 11, 13, 13, 13, 16, 17, 18, 18, 19, 20, 22, 23, 27, 28, 31, 32, 34, 37, 37, 41, 44, 44, 47, 49};
static int    Gj[NNZ]   = {25, 13, 1, 5, 7, 35, 18, 19, 31, 32, 21, 32, 33, 0, 8, 27, 16, 25, 3, 24, 17, 27, 13, 3, 28, 29, 30, 2, 23, 29, 31, 10, 8, 29, 1, 20, 22, 3, 8, 16, 19, 10, 14, 24, 2, 6, 15, 17, 34};
static double Ga[NNZ]   = {8, 3, 7, 5, 6, 7, 1, 9, 8, 9, 9, 9, 5, 1, 5, 8, 4, 4, 2, 11, 3, 8, 9, 7, 7, 4, 2, 2, 7, 6, 9, 3, 4, 2, 2, 9, 7, 4, 7, 8, 1, 8, 6, 1, 3, 3, 6, 6, 1};
static double Xa[N]     = {3, 2, 2, 7, 1, 5, 3, 3, 6, 6, 4, 8, 8, 4, 7, 8, 9, 7, 7, 6, 9, 5, 8, 5, 7, 5, 5, 5, 2, 4, 8, 1, 3, 6, 9, 8};
static double Za[M]     = {40, 0, 12, 113, 69, 27, 0, 45, 0, 57, 0, 0, 73, 36, 20, 0, 14, 77, 61, 36, 95, 4, 68, 12, 32, 141, 0, 148, 81, 0, 63, 51};
static double Ya[M];

// Petsc MATMPIAIJ matrix
typedef struct {
  int     m, n;   // local row/column size
  int     M, N;   // global row/cloum size
  int    *i;      // row pointer
  int    *j;      // column indices
  double *a;      // values
  int     rstart; // start row index on this process
  int     cstart; // start column index on this process
} Mat;

// Petsc VECMPI vector
typedef struct {
  int     n, N; // local/global size
  double *a;    // values
} Vec;

int main(int argc, char **argv)
{
  Mat    A; // MPI parallel representation of the matrix
  Vec    x, y, z;
  double norm = 0;
  int    rank, size;
  int    m; // Number of rows on the current MPI process
  int    n; // Number of columns on the current MPI process

  MPI_Init(&argc, &argv);
  MPI_Comm_rank(MPI_COMM_WORLD, &rank);
  MPI_Comm_size(MPI_COMM_WORLD, &size);

  // Distribute A by rows
  m = M / size + ((M % size) > rank ? 1 : 0); // Actually in petsc one can set m arbitrarily, as long as SUM(m) = M
  n = N / size + ((N % size) > rank ? 1 : 0); // Actually in petsc one can set n arbitrarily, as long as SUM(n) = N

  // Compute A row layout
  MPI_Exscan(&m, &A.rstart, 1, MPI_INT, MPI_SUM, MPI_COMM_WORLD);
  if (rank == 0) A.rstart = 0; // otherwise undefined by MPI_Exscan

  // Compute A column layout. Columns are not distributed, but have a layout conforming to the distribution of x in y = Ax
  MPI_Exscan(&n, &A.cstart, 1, MPI_INT, MPI_SUM, MPI_COMM_WORLD);
  if (rank == 0) A.cstart = 0;

  A.m = m;
  A.M = M;
  A.n = n;
  A.N = N;

  A.i    = (int *)malloc(sizeof(int) * (m + 1));
  A.i[0] = 0;
  for (int i = 0; i < m; i++) A.i[i + 1] = A.i[i] + (Gi[A.rstart + i + 1] - Gi[A.rstart + i]);

  A.j = Gj + Gi[A.rstart]; // Can directly reuse Gj, Ga, but not Gi
  A.a = Ga + Gi[A.rstart];

  // Distribute x according to A's column partition
  x.n = A.n;
  x.N = A.N;
  x.a = Xa + A.cstart;

  // Distribute y, z according to A's row distribution
  y.n = A.m;
  y.N = A.M;
  y.a = Ya + A.rstart;

  z.n = A.m;
  z.N = A.M;
  z.a = Za + A.rstart;

  // All gather the distributed x to a local X[]. We don't use Xa[] because we want to simulate real code where Xa[] is not available
  double *X = (double *)malloc(sizeof(double) * x.N);

  if (N % size) {
    int *recvcounts = (int *)malloc(sizeof(int) * size);
    int *disp       = (int *)malloc(sizeof(int) * size);

    MPI_Allgather(&n, 1, MPI_INT, recvcounts, 1, MPI_INT, MPI_COMM_WORLD);

    disp[0] = 0;
    for (int i = 1; i < size; i++) disp[i] = disp[i - 1] + recvcounts[i - 1];

    MPI_Allgatherv(x.a, x.n, MPI_DOUBLE, X, recvcounts, disp, MPI_DOUBLE, MPI_COMM_WORLD);
    free(recvcounts);
    free(disp);
  } else { // A columns are distributed evenly
    MPI_Allgather(x.a, x.n, MPI_DOUBLE, X, x.n, MPI_DOUBLE, MPI_COMM_WORLD);
  }

  // Compute y = AX
  for (int i = 0; i < A.m; i++) {
    y.a[i] = 0.0;
    for (int j = A.i[i]; j < A.i[i + 1]; j++) y.a[i] += A.a[j] * X[A.j[j]];
  }

  // Computing the norm of ||y - z||
  for (int i = 0; i < A.m; i++) norm += (y.a[i] - z.a[i]) * (y.a[i] - z.a[i]);
  MPI_Allreduce(MPI_IN_PLACE, &norm, 1, MPI_DOUBLE, MPI_SUM, MPI_COMM_WORLD);

  free(A.i);
  free(X);

  MPI_Finalize();
  if (rank == 0) {
    if (norm > 1e-6) {
      printf("Error in computing y = Ax, with norm = %f\n", norm);
      return 1;
    } else {
      printf("Succeeded in computing y = Ax\n");
      return 0;
    }
  }
}
\end{lstlisting}

\end{document}